	\documentstyle[12pt,group21]{article}
\title{A general algebraic model for vibrational  molecular
spectroscopy}
\author{ A. Frank\adr{1,2} R. Lemus\adr{1} F. P\'erez-Bernal\adr{3}
R. Bijker\adr{1} \and \ J.M. Arias\adr{3}}

\address[1]{Instituto de Ciencias Nucleares, U.N.A.M.,\\
         A.P. 70-543, 04510 M\'exico D.F., M\'exico}
\address[2]{Instituto de F\'{\i}sica, Laboratorio de Cuernavaca,\\
         A.P. 139-B, Cuernavaca, Morelos, M\'exico}
\address[3]{Departamento de F\'{\i}sica At\'omica, Molecular y Nuclear,\\
         Facultad de F\'{\i}sica, 
         Apdo. 1065, 41080 Sevilla, Espa\~na}
\newcommand{\ba}{\begin{eqnarray}}
\newcommand{\ea}{\end{eqnarray}}
\begin{document}
\maketitle

\section{Algebraic Model}
We present a symmetry-adapted version of the vibron model\cite{A1}.
The model exploits the isomorphism of the $U(2)$ Lie Algebra and the one 
dimensional Morse oscillator\cite{E}. A $U(2)$ algebra is assigned to each 
relevant interatomic interaction. The operators in the model are 
expressed in terms of the generators of the molecular dynamical group\cite{B},
which in the case of triatomic molecules is given by the product
\ba
U^{1} (2) \otimes U^{2} (2) \otimes U^{3} (2) ~~ . \label{dingr} 
\ea

A simple realization for those generators is given in terms of angular momentum
$\hat J_{\nu,i}$ and number $\hat N_i$ operators
\ba
\{ \hat N_i , \hat J_{x,i}, \hat J_{y, i} , \hat J_{z,i} \} , \quad i
= 1, 2, 3~~. \label{geners} 
\ea
Instead of working with the generators in Eq.~(\ref{geners}) we introduce a 
new set of 
generators with well-defined tensorial properties under the point group\cite{A}.
The choice of cartesian coordinate system, irreducible representations and 
Clebsch-Gordan coefficients is given in Ref.\cite{B}.

The use of the symmetry-adapted generators allows the connection between the 
algebraic and the configuration space calculations clarifying the geometrical 
content of the algebraic approach\cite{A}.


	The relevant symmetry projected generators for {\em D}$_{3h}$ 
triatomic molecules are 
\ba
\hat T^{A_1}_{\mu, 1} = {1\over \sqrt{3}} \left( \hat
      J_{\mu, 1} + \hat J_{\mu, 2} + \hat J_{\mu, 3} \right) ~~&,&  
\nonumber \\   
\hat T^E_{\mu, 1} = {1\over \sqrt{6}} \left( 2 \hat J_{\mu, 1} -
             \hat J_{\mu, 2}  - \hat J_{\mu, 3} \right) ~ &,&
\hat T^E_{\mu, 2} = {1\over \sqrt{2}} \left( \hat J_{\mu, 2} - 
                      \hat J_{\mu, 3} \right) ~ ,\label{symmgen}
\ea
with $\mu = +, -, 0$. 

The algebraic Hamiltonian is constructed by repeated couplings 
of these tensors to a total symmetry $A_1$. Terms quadratic in the generators 
and its products can be expressed in terms of Casimir operators used in former 
algebraic approaches\cite{E,F}. However, other couplings, which are 
physically relevant, like $\hat l^2$ written below cannot be expressed in terms 
of Casimir operators and consequently are not in previous algebraic approaches.

According to this we obtain the Hamiltonian
\ba
\hat {\cal H}  = \alpha \hat {\cal H}_{A_1} + \beta \hat{\cal
H}_E + \gamma \hat {\cal V}_{A_1} + \delta \hat l^2 + \alpha^{[2]} \hat
{\cal H}^{2}_{A_1} + \beta^{[2]} \hat{\cal H}^2_E + \xi^{[2]} \hat {\cal
H}_{A_1E} + \epsilon (\hat T^3_+ + T^3_- ) ~~ , \label{hamil}
\ea
where
\ba
\hat{\cal H}_{\Gamma_x} = \frac{1}{2N_{x}} \sum_{\gamma} \left( 
  \hat T^{\Gamma_x}_{-,\gamma} \, \hat T^{\Gamma_x}_{+,\gamma}
+ \hat T^{\Gamma_x}_{+,\gamma} \, \hat T^{\Gamma_x}_{-,\gamma} 
\right) ~,~~
\hat{\cal V}_{\Gamma_x} = \frac{1}{N_{x}} \sum_{\gamma} \,
\hat T^{\Gamma_x}_{0,\gamma} \, \hat T^{\Gamma_x}_{0,\gamma} ~,
\label{hv}
\ea
and
\ba 
\hat l = -i \, \sqrt{2} \frac{1}{N_b} 
[ \hat T^{E}_{-} \times \hat T^{E}_{+} ]^{A_2} ~&,&~~\hat T_\pm = 
\hat T^E_1 \pm i \hat T^E_2~, 
\nonumber\\
\hat{\cal H}_{A_1E}\,  \equiv  \,  {(\hat{\cal H}_{A_1}\hat{\cal H}_E
+ \hat{\cal H}_{E} \hat {\cal H}_{A_1}) \over 2}&.&
\label{gij}
\ea
The Hamiltonian diagonalization and parameter fitting procedures are enhanced 
using a symmetry-adapted basis\cite{B}.

\section{ Application to H$^+_3$, Be$_3$ and Na$^+_3$ }

The three  {\em D}$_{3h}$ symmetric chosen molecules exhibit a wide range of 
behaviours, from the very 
anharmonic spectrum of H$_3^+$ to the almost harmonic of Na$_3^+$.

We present in Table I a least square fit calculation to the {\em ab initio} 
calculated  vibrational spectrum for these molecules, 
using an optimal set of parameters of the Hamiltonian. In the H$_3^+$ case, 
due to its anharmonicity, we had to include the full set of interactions 
in Eq.\ (\ref{hamil}) to reproduce the spectrum. 
The relationship between operators and anharmonicity is clear once the 
harmonic limit of the AOSM model is explored\cite{A} and the link to 
configuration space calculations is analyzed.

In addition to the $8$ parameters in Eq.\ (\ref{hamil}), the value of the 
boson number $N$\cite{E} has to be fixed and it was taken to be $N_{Na_3^+} = 
N_{Be_3} = N_{H_3^+} = 30$.

This work suggests that the AOSM represents a systematic, simple and accurate
alternative to configuration space methods when the integro-differential 
approach becomes too complex to be applied.
\begin{table}[t]
\centering
\caption[]{\small Least square energy fit to {\em ab initio} 
calculations\cite{G} for Na$^+_3$, Be$_3$ and H$^+_3$  using 
Hamiltonian (\ref{hamil}).  We show
 the energy differences $\Delta E = E_{ab init} - E_{alg}$.  
All energies are given in $cm^{-1}$.} 
\begin{tabular}{|c|c|c|c|c|}
\hline 
 & & Na$^+_3$ & Be$_3$ & H$^+_3$ \\
(v$_{A_1}$ v$^l_E$) & $\Gamma$ & $\Delta$E & $\Delta$E & $\Delta$E \\ 
\hline
(01$^1$) & e   &   0.93    &   0.51  & -1.55  \\ 
(10$^0$) & a$_1$ &   1.95    &   0.02  &  0.42  \\ 
(02$^0$) & a$_1$ &   0.37    &   -0.74 &  7.48  \\ 
(02$^2$) & e   &   0.84    &   0.17  &  -5.69 \\ 
(11$^1$) & e   &   1.68    &   0.82  &  -0.61 \\ 
(20$^0$) & a$_1$ &   1.26    &   -0.04 &  -0.11 \\  
(03$^1$) & e   &   -1.19   &   -2.05 &  -4.46 \\ 
(03$^3$) & a$_1$ &   -0.34   &   -1.23 &  3.18  \\ 
(03$^3$) & a$_2$ &   -0.33   &   0.61  &  2.44  \\ 
(12$^0$) & a$_1$ &   -0.01   &   1.90  &  0.66  \\ 
(12$^2$) & e   &    0.34   &   -1.36 &  -5.00 \\ 
(21$^1$) & e   &   -0.19   &   0.79  &  4.07  \\ 
(30$^0$) & a$_1$ &   -2.06   &   -1.66 &  -1.23 \\ 
\hline
 &  rms & 1.33 & 1.24 & 5.84\\
\hline
         & $\alpha$ & 142.40  &  458.91   & 3193.60 \vline $\,
\alpha^{[2]}$ -14.86 \\ 
Parameters & $\beta$  & 100.32  &  396.27 & 2507.16 \vline $\,
\beta^{[2]}$ -27.75 \\
           & $\gamma$ &  21.31  &  209.74 & 2807.83$ \,$\vline $\,
\xi^{[2]}$ -28.04 \\ 
           & $\delta$ & -0.19  &  -0.95 &  -13.44 \vline $\,
\epsilon\;\;\;$ -0.90 \\ 
\hline
\end{tabular}
\label{tabla}
\end{table}

Support by the European Community, contract nr. CI1$^{\ast}$-CT94-0072, 
DGAPA-UNAM, project IN105194, CONACyT-M\'exico, project 
400340-5-3401E and Spanish DGCYT, project PB95-0533, is acknwoledged.

\end{document}